# Two body non-leptonic decays of $\Lambda_b$

**Gurpreet Sidana,**

**Department of Physics,**

**Guru Nanak Dev University,**

**Amritsar143005, India**

**Abstract**

Two body non-leptonic decays: $\Lambda_b \to \Lambda_c P\ (V)$ are investigated. Baryon form factors of the heavy quark effective theory are employed. Form factors for renormalization in full QCD are also considered. We find that contributions of the non-factorizable terms are negligible in b-decays. Comparison with calculations in other models shows that form factors for renormalization in full QCD should be more relevant. The asymmetry parameters are negative as in the charmed baryons cases. An experimental measure on the form factors $V_1$ and $A_1$ is accentuated to test the validity of HQET.

**1. Introduction**

The studies of the b-quark are an interesting probe for measuring and understanding some of the fundamental and delicate parameters of the Standard Model– origin of electroweak mixing, quark masses, CP-violation, and its proposed interpretation in terms of electroweak mixing. In addition, b-decays are expected to be dominated by the short distance dynamics thereby providing an important testing ground for perturbative QCD. However most of the theoretical effort has gone in investigating B-meson decays. Advancement in the arena of heavy baryons both experimental and theoretical has been very slow [1-6].

From theoretical point of view, the dynamics of non-leptonic weak decays of hadrons is expected to become simpler as the quark becomes heavier. As in the case of mesons, all hadronic weak decays of baryons can be expressed in terms of the following quark diagrams amplitudes [7]: (a) the external W-emission diagram; (b) the internal W-emission diagram (c) the W-exchange diagram (c) the horizontal W-loop diagram (penguins). The external and internal W- emission diagrams are referred to as the color-allowed and color-supressed contributions. However, baryons being made of three quarks in contrast to two quarks for mesons bring along several complications.

First of all, the non-factorizable terms such as W-exchange contributions are negligible relative to the factorizable ones are known empirically to be working reasonably well for describing the non-leptonic weak decays of heavy mesons [8]. Only for the $B \to \pi\pi, k\pi$ modes the addition of penguins is required to have agreement with the experimental data [9]. However this approximation is a priori not applicable to baryons as w-exchange terms, manifested as pole diagrams, are no longer subject to helicity and color suppressions. Therefore the pole contributions is as important as the factorizable one [10]. This has been proved so by the experimental measurement of the decay modes $\Lambda_c^+ \to \Sigma^0 \pi^+, \Sigma^+ \pi^0, \Xi^0 k^+$, which do not receive any factorizable contribution. The W-exchange term therefore, plays an essential role in charmed baryon decays. Hence their contribution in bottom baryon decays may be equally significant.

A renewed motivation for the study of heavy hadrons came with the formulation of an effective field theory where the heavy quark mass goes to infinity, with its four-velocity fixed. The dynamics of the atom-like hadrons then experiences great simplification due to the appearance of a new $SU(2) \otimes SU(2)$ spin-flavor symmetry [11]. The hadronic form factors are then expressed in terms of a single universal function $\zeta(\upsilon . \upsilon')$, the Isgur-Wise (IW) function, which depends only on the four-velocities $\upsilon_m$ of heavy particles and is renormalized at zero recoil ($\tilde{\upsilon}\ \upsilon'$ = 1) [12]. Away from the symmetry limit, these predictions of HQET receive corrections arising from finite quark masses [13] and hard gluon exchange. The perturbative QCD corrections are calculated both in the effective theory and full QCD [14,15].

Encouraged by the consistency between experiment and non-factorizable contributions for the charm baryon decays in our earlier work [16], we would like to present this work a systematic study of exclusive two body decays of bottom baryons. The modes $\Lambda_b \to \Lambda_c P\ (V)$ are considered in the framework of soft-meson techniques with the inclusion of factorizable terms. Form factors of HQET are employed. We find that contribution of current algebra terms in $\Lambda_b$ decays is quiet suppressed. Taking factorization as the dominant contribution we then incorporate the perturbative QCD effects in full QCD. Since the decay widths are functions only of the Isgur-Wise form factor, their study may provide an insight into the long-distance dynamics involved in hadronic weak decays.

Section II describes the weak Hamiltonian and weak decay amplitudes for renormalization in the effective theory as well as full QCD. Details of numerical calculations are given in section III, followed by discussion of results in section IV. We summarize our conclusions in section V.

**2. Mathematical Framework**

**2.1 Weak Hamiltonian**

For the two-body non-leptonic decay of the type

$$B_b(p) = B_c(p) + M(q)$$

where $B_b$ is a baryon that contains one b-quark while $B_c$ is a baryon containing c-quark. $M$ is a meson with spin parity $J^P = 0^-(P);1^-(V)$, the effective weak Hamiltonian including the short-distance QCD effects for the b$\rightarrow$c transition is as follows:

**CKM-allowed**:

$$H_W^{eff} = \frac{G_F}{\sqrt{2}} V_{bc}\{c_1[(\bar{c}b)(\bar{d}u)V_{ud} + (\bar{c}b)(\bar{s}c)V_{cs}] + c_2[(\bar{d}b)(\bar{c}u)V_{ud} + (\bar{s}b)(\bar{c}c)V_{cs}]\} \tag{1}$$

**CKM- suppressed**:

$$H_W^{eff} = \frac{G_F}{\sqrt{2}} V_{bc}\{c_1[(\bar{c}b)(\bar{s}u)V_{us} + (\bar{c}b)(\bar{d}c)V_{cd}] + c_2[(\bar{s}b)(\bar{c}u)V_{us} + (\bar{d}b)(\bar{c}c)V_{cd}]\} \tag{2}$$

where $\bar{q}q = \bar{q}\gamma_\mu(1-\gamma_5)q,\ c_1(m_b) \cong 1.1$ and $c_2(m_b) \cong -0.25.$

**2.2. Weak Decay Amplitudes (Renormalization in effective theory)**

There have been theoretical attempts to analyze non-leptonic heavy baryon decays using factorizing contribution alone [17,18], the argument being that the W-exchange (pole-terms) contributions can be neglected in analogy with power-suppressed W-exchange contributions in the inclusive non-leptonic decays of heavy baryons. One might even be attempted to drop the non- factorizing contributions on account of the fact that they are superficially proportional to $1/N_c$. However, since $N_c$ baryons contain $N_c$ quarks an extra combinational factor proportional to $N_c$ appears in the amplitudes which cancels the explicit diagrammatic $1/N_c$ factor [19]. There is now ample empirical evidence in the charm sector that the non-factorizing diagrams can not be neglected [20]. Consequently we proceed to analyse bottom baryons decays with the inclusion of non-factorizable terms.

### 2.2.1 $\Lambda_b^o \to \Lambda_c^+ P$ Decays

Applying the standard current algebra techniques, the evaluation of $\Lambda_b \to \Lambda_c P$ involves relating the three hadron amplitude $<\Lambda_C P|H_W|\Lambda_b>$ to the baryon-baryon transition matrix element $<\Lambda_C|H_W|\Lambda_b>$ through PCAC. Adding the factorization term gives the general form:

$$<\Lambda_C P_k|H_W|\Lambda_b> = <\Lambda_C|[Q_k^5, H_W]|\Lambda_b> + M_{pole} + M_{fac} \tag{3}$$

The first term corresponds to the equal-time commutator, which is essentially the matrix element of $H_w$ between the two ½$^+$ baryon states:

$$<\Lambda_C|H_w^{eff}|\Lambda_b> = \bar{h}_c(a + b\gamma_5)h_b \tag{4}$$

since $<B_f|H_w^{eff}|B_i> = 0$, the ETC contributes only to the s-wave amplitude:

$$A^{ETC} = \frac{1}{f_p} <\Lambda_C|[Q_k^5, H_w^{pv}]|\Lambda_b> = \frac{1}{f_p} <\Lambda_C|[Q_k, H_w^{pc}]|\Lambda_b> \tag{5}$$

$Q_k$ and $Q_k^5$ are the vector and coaxial-vector charges respectively. The p-waves are then described by the ½$^+$ pole contributions. The baryon pole terms arising from the s- and u-channels contribute only to the PC amplitude and are given by:

$$B^{pole} = \frac{g_{lfk} a_{il}}{m_i - m_l} \frac{m_i + m_f}{m_l + m_f} + \frac{g_{il'k} a_{l'f}}{m_f - m_{l'}} \frac{m_i + m_f}{m_i + m_{l'}} \tag{6}$$

where $g_{ijk}$ are the strong baryon-meson coupling constants, and *l, l'* are the intermediate baryon states corresponding to the respective channels.

The third term is the factorization term obtained by inserting vacuum intermediate states which reduces it to a product of two-current matrix elements that vanish in the soft-meson limit. The separable combination for $\Lambda_b \to \Lambda_c P$ is:

$$<P_k|J_\mu|0><\Lambda_c|J_\mu|\Lambda_b> = f_p p^\mu H_\mu \qquad (7)$$

where

$$<P_k|J_\mu|0>= f_p p^\mu \qquad (8)$$

$p^\mu$ is the four-momentum of the pseudoscalar meson and $f_P$ is its decay constant. The hadronic current describing the baryon-baryon transition matrix element is defined in terms of the invariant vector and axial-vector form factors as

$$\begin{aligned} H_\mu &= <\Lambda_c|J_\mu|\Lambda_b> \\ &= h_c(\upsilon')[(F_1\gamma_\mu + F_2\upsilon_\mu + F_3\upsilon'_\mu) - (G_1\gamma_\mu + G_2\upsilon_\mu + G_3\upsilon'_\mu)\gamma_5]h_b(\upsilon) \end{aligned} \qquad (9)$$

where $F_i$ and $G_i$ are the vector and axial-vector form factors respectively and are functions of $\upsilon.\upsilon'$ *(=y* hereafter).

As a first approximation, we use HQET predictions for the baryon form factors. In the limit of infinite bottom and charm quark masses there is only one independent form factor in the $\Lambda_b^o \to \Lambda_c^+$ current matrix element [13].

$$\begin{aligned} F_1(y) &= G_1(y) = \varsigma(y) \\ F_2(y) &= G_2(y) = F_3(y) = G_3(y) = 0 \end{aligned} \qquad (10)$$

where

$$\varsigma(y) = \left(\frac{\alpha_s(m_b)}{\alpha_s(m_c)}\right)^{-6/25}\left(\frac{\alpha_s(m_c)}{\alpha_s(m_\mu)}\right)^{a_L(y)} \varsigma_o(y) \qquad (11)$$

incorporates the radiative corrections. y = $\upsilon.\upsilon'$ is the velocity transfer between the initial and final baryon states. The velocity-dependent anomalous dimension $a_L(y)$ is given by

$$a_L(y) = \frac{8}{27}\left\{\frac{y}{\sqrt{y^2-1}}\ln(y+\sqrt{y^2-1})-1\right\} \tag{12}$$

μ is the typical hadronic scale chosen at the charm quark mass in our computations. The function $\zeta_o(y)$ is the non-perturbative contribution and contains information about the long-distance dynamics.

The factorized PV and PC amplitudes are now given as:

$$A^{fac} = c_k \frac{G_F}{\sqrt{2}} V_{cj} V_{kl}^* f_p (m_{\Lambda_b} - m_{\Lambda_c}) \varsigma(y)$$

$$B^{fac} = c_k \frac{G_F}{\sqrt{2}} V_{cj} V_{kl}^* f_p (m_{\Lambda_b} + m_{\Lambda_c}) \varsigma(y) \tag{13}$$

$\varsigma(y)$ is defined by (11) and incorporates the perturbative QCD effects in the effective theory. (c,k) and (j,l) refers to the +2/3 charge quarks and -1/3 charge quarks respectively. $c_k$ is the QCD coefficient equal to $c_1(c_2)$ depending on the emitted meson.

The total amplitude is now given by

$$A = A^{BTC} + A^{fac}$$
$$B = B^{pole} + B^{fac} \tag{14}$$

The associated decay widths Γ and asymmetry parameter α are computed as

$$\Gamma(\Lambda_b^o \to \Lambda_c^+ + P^-) = \frac{|\vec{p}_{\Lambda_c}|}{4\pi m\Lambda_b}(E_{\Lambda_C} + m_{\Lambda_c})[|S^2| + |P^2|] \tag{15}$$

where

$$S = A$$

$$P = B\frac{|\vec{p}_{\Lambda_c}|}{E_{\Lambda_C} + m_{\Lambda_c}} \tag{16}$$

and the symmetry parameter

$$\alpha = \frac{2\operatorname{Re} S^* P}{|S|^2 + |P|^2} \tag{17}$$

**2.2.2 $\Lambda_b^o \to \Lambda_c^+ + V^-$ Decays**

For the emission of a vector meson V in the weak decays of $\Lambda_b$, the matrix element follows in analogy with the $\Lambda_b \rightarrow \Lambda_c P$ decays. However, we find from numerical calculations and the work of other peoples [2,19] that the non-fectorizable contribution in bottom baryon decays are small. Taking factorization to be the dominant contribution, the factorized amplitude is given by

$$M_{\Lambda_b^o \rightarrow \Lambda_c^+ V^-} = f_V H_\mu \varepsilon^\mu \tag{18}$$

in analogy to eqn. (7). $\varepsilon^\mu$ is the polarization vector of the vector meson and $f_V$ is its decay constant. Substituting for $H_\mu$ from (9) and using the HQET limit the relevant matrix element becomes:

$$M_{\Lambda_b^o \rightarrow \Lambda_c^+ V^-} = \frac{G_F}{\sqrt{2}} V_{cj} V_{kl}^* f_V \varsigma(y) \bar{h}_c(\upsilon')[\gamma_\mu(1-\gamma_5)] h_b(v) \varepsilon^\mu \tag{19}$$

$V_{cj}$ and $V_{kl}$ are the appropriate CKM matrix elements. The product of the phase space factor and the square of the amplitude give the decay width:

$$\Gamma(\Lambda_b^o \rightarrow \Lambda_c^+ V^-) = \frac{\left|\vec{p}_{\Lambda_c}\right|}{16\pi m_{\Lambda_b}} m_{\Lambda_c} \left|M\right|^2 = \frac{G_F^2}{4\pi m_{\Lambda_b}} \left|V_{cj}\right|^2 \left|V_{kl}\right|^2 f_V^2 \varsigma(y)^2 \left|\vec{p}_{\Lambda_c}\right| m_{\Lambda_c} (E_V x_V + m_V^2 y) \tag{20}$$

where

$$\left|\vec{p}_{\Lambda_c}\right| = \frac{1}{2m_{\Lambda_b}} \sqrt{[m_{\Lambda_b}^2 - (m_{\Lambda_c} - m_V)^2][m_{\Lambda_b}^2 - (m_{\Lambda_c} + m_V)^2)}$$

$$E_V = \frac{m_{\Lambda_b}^2 - m_{\Lambda_c}^2 + m_V^2}{2m_{\Lambda_c}}$$

$$x_V = \frac{m_{\Lambda_b}^2 - m_{\Lambda_c}^2 + m_V^2}{m_{\Lambda_c}} \tag{21}$$

**2.3. Renormalization in full QCD**

The baryon matrix element in the full theory can be expressed in terms of the heavy quark form factors as:

$$H_\mu = \varsigma_O(y) \bar{h}_c(\upsilon')[(V_1\gamma_\mu + V_2\upsilon_\mu + V_3\upsilon'_\mu) - (A_1\gamma_\mu + A_2\upsilon_\mu + A_3\upsilon'_\mu)\gamma_5] h_b(\upsilon) \tag{22}$$

where $\varsigma_o(y)$ was used in (11). Note that the pertabative QCD corrections are now computed in full QCD and are not just the multiplicative term as in eqn. (11). The complete order-$\alpha_s$ renormalization of the heavy quark operators including the RG-improvement leads to following expression for the heavy quark form factors $V_i$ and $A_i$ [15]:

$$V_1 = Z_{1R}(y,\lambda)\left(\frac{\alpha_s(m_b)}{\alpha_s(m_c)}\right)^{-6/25}\left[1+\frac{\alpha_s(\bar{m})}{\pi}\tilde{\upsilon}_1(y)\right]$$

$$A_1 = Z_{1R}(y,\lambda)\left(\frac{\alpha_s(m_b)}{\alpha_s(m_c)}\right)^{-6/25}\left[1+\frac{\alpha_s(\bar{m})}{\pi}\tilde{a}_1(y)\right]$$

$$V_{2,3} = Z_{1R}(y,\lambda)\left(\frac{\alpha_s(m_b)}{\alpha_s(m_c)}\right)^{-6/25}\frac{\alpha_s(\bar{m})}{\pi}\tilde{\upsilon}_{2,3}(y)$$

$$A_{2,3} = Z_{1R}(y,\lambda)\left(\frac{\alpha_s(m_b)}{\alpha_s(m_c)}\right)^{-6/25}\frac{\alpha_s(\bar{m})}{\pi}\tilde{a}_{2,3}(y) \tag{23}$$

with

$$Z_{1R} = 1-\frac{2\alpha_s}{3\pi}[yr(y)-1]\ln\frac{m_c^2}{\lambda}, \qquad Z(1,\lambda)=1 \tag{24}$$

The baryon matrix element (22) then becomes:

$$H_\mu = \varsigma_o(y)Z_{1R}(y,\lambda)\left(\frac{\alpha_s(m_b)}{\alpha_s(m_c)}\right)^{-6/25}.\ \bar{h}_c(\upsilon')\{[1+\frac{\alpha_s(\bar{m})}{\pi}(\tilde{\upsilon}_1\gamma_\mu+\tilde{\upsilon}_2\upsilon_\mu+\tilde{\upsilon}_3\upsilon'_\mu)]$$

$$+[1+\frac{\alpha_s(\bar{m})}{\pi}(\tilde{a}_1\gamma_\mu+\tilde{a}_2\upsilon_\mu+\tilde{a}_3\upsilon'_\mu)\gamma_5]\}h_b(\upsilon) \tag{25}$$

The RG- improved matrix element for the decay $\Lambda_b^o \to \Lambda_c^+ P^-$ is

$$M_{\Lambda_b^o\to\Lambda_c^+P^-} = \frac{G_F}{\sqrt{2}}V_{cj}V_{kl}^* f_p \varsigma_o(y)\bar{h}_c(\upsilon')\,[(V_1\gamma_\mu+V_2\upsilon_\mu+V_3\upsilon'_\mu)$$

$$-(A_1\gamma_\mu+A_2\upsilon_\mu+A_3\upsilon'_\mu)\gamma_5]h_b(\upsilon) \tag{26}$$

$V_i$ and $A_i$ are defined by (23). The associated decay width has the form:

$$\Gamma(\Lambda_b^o \to \Lambda_c^+ P^-) = \frac{G_F^2}{4\pi m_{\Lambda_b}} |V_{cj}|^2 |V_{kl}|^2 f_p^2 |\varsigma(y)|^2 \; |\vec{p}_{\Lambda_c}| m_{\Lambda_c}$$

$$\{4V_1^2[E_p x_p + m_p^2(1-y)] + 4A_1^2[E_p x_p - m_p^2(1+y)]$$
$$+ 4V_1V_2[E_p(x_p + 2E_p)] - 4A_1A_2[E_p(x_p - 2E_p)]$$
$$+ 2V_1V_3[x_p(2E_p + x_p)] + 2A_1A_3[x_p(2E_p - x_p)]$$
$$+ 4V_2V_3[x_p E_p(1+y)] - 4A_2A_3[x_p E_p(1-y)]$$
$$+ 4V_2^2[E_p^2(1+y)] - 4A_2^2[E_p^2(1-y)]$$
$$+ V_3^2[x_p^2(1+y)] - A_3^2[x_p^2(1-y)] \qquad (27)$$

where $x_p$ and $E_p$ are analogously defined by eqn. (21).

Similarly for the vector meson case, the expression for the decay width is

$$\Gamma(\Lambda_b^o \to \Lambda_c^+ V^-) = \frac{G_F^2}{4\pi m_{\Lambda_b}} |V_{cj}|^2 |V_{kl}|^2 f_V^2 |\varsigma(y)|^2 \; |\vec{p}_{\Lambda_c}| m_{\Lambda_c}$$

$$\{4V_1^2[E_V x_V + m_V^2(3-y)] + 4A_1^2[E_V x_V + m_V^2(3+y)]$$
$$+ 4V_1V_2[E_V(2E_V + x_V) - 2m_V^2(1+y)]$$
$$+ 4A_1A_2[E_V(2E_V - x_V) - 2m_V^2(1-y)]$$
$$+ 2V_1V_3[x_V(2E_V + x_V) - 4m_V^2(1+y)]$$
$$+ 2A_1A_3[x_V(2E_V - x_V) + 4m_V^2(1-y)] \; + 4V_2V_3(1+y)[x_V E_V - 2ym_V^2]$$
$$- 4A_2A_3[x_V E_V(1-y) + 2ym_V^2(1+y)]$$
$$+ 4V_2^2(1+y)[E_V^2 - m_V^2] - 4A_2^2(1-y)[E_V^2 - m_V^2]$$
$$+ V_3^2(1+y)[x_V^2 - 4m_V^2] - A_3^2(1-y)[x_V^2 - 4m_V^2]\} \qquad (28)$$

**3. Numerical Estimates**

For numerical values of the baryon and meson masses and the KM-matrix elements, we refer to the Particle Data Group, 2004 [21]. The quark masses are set to be $m_c$ = 1.5 GeV   $m_b$ = 4.9 GeV respectively and the parameter $\bar{\Lambda}$ = 0.75 GeV. The universal form factor $\zeta_o(y)$ is taken to be in the exponential form:

$$\zeta_o = \exp b(1-y) \tag{29}$$

where *b* is a constant to be determined by experimental data. Ito et al.[22] estimate this value to be 1.085 using the experimental data for $\overline{B}^o \to D^* l \overline{\nu}_l$. Assuming the same value holds for baryon decays, we make predictions for the absolute decay rates. The vector mesons decay constant values are taken as $f_\rho = f_{k*} = 0.221 GeV$, as determined by the experimental data on $\rho ? e^+ e^-$ and $\tau ?\ \nu_\tau \rho\ \rho$. The IW function is evaluated at the point

$$\upsilon.\upsilon' = y = \frac{m_{\Lambda_b}{}^2 + m_{\Lambda_c}{}^2 - m_p{}^2}{2 m_{\Lambda_b} m_{\Lambda_c}} \cong 1.45 GeV \tag{30}$$

for the pion and K-mesons modes and at y = 1.3 GeV for the $D^-$ and $D_s^-$ mesons. The vector mesons modes have y = 1.4 GeV.

For estimation of the pole term we have related the weak matrix elements eg. $a_{\Lambda_b^o \Sigma_c^o}$ by *SU(5)* symmetry to the matrix element $a_{\Sigma^+ p}$ through the relation

$$a_{\Lambda_b^o \Sigma_c^o} = \frac{1}{\sqrt{6}} \frac{V_{bc}}{V_{us}} < p \left| H_W^{pc} \right| \Sigma^+ > \tag{31}$$

The numerical value of $a_{\Sigma^+ p}$ is taken from [23]. Since *SU(5)* symmetry is very badly broken, we estimate the broken coupling constants employing the Coleman-Glashow null result for tadpole-type symmetry breaking. The baryon-meson coupling constants [24] are then given by

$$g_{BB'P} = \frac{M_B + M_{B'}}{2 M_N} \frac{1}{\alpha_P} g_{BB'P}^{sym} \tag{32}$$

where $g_{BB'P}^{sym}$ is the SU(5) predicted value. $\alpha_p$=1 for π, K- mesons and $\alpha_P = \frac{\delta M_c}{\delta M_s}$ for D-mesons.

The numerical values of the matrix elements in the ETC term are the same as for the pole term contributions. The computed values of the weak amplitude, decay widths and asymmetry parameter are listed in Tables 1 to 3.

**4. Results And Discussion**

The results of Table 1 show that the Cabbibo-allowed modes $\Lambda_b^o \to \Lambda_c^+ \pi^-$, $\Lambda_b^o \to \Lambda_c^+ D_S^-$ form a considerable fraction of the total width of $\Lambda_b$, the other two being relatively suppressed. The ratios $|A_{pole}/A_{fac}| \simeq .16$ and $|B_{pole}/B_{fac}| \simeq .21$ imply that the contributions of the current algebra terms is largely suppressed in the bottom baryon sector. The results agree with Xu and Kamal [2] who find the W-exchange amplitude to be ~6% of the total amplitude. The up-down asymmetry parameter $\alpha \sim -1$ for all modes. The results compare with Cheng [3] who also find the parameter $\alpha$ to be negative. This implies that one should expect maximal asymmetry in the angular distribution of Λ's even in the heavier sector.

For the vector meson modes we find the branching ratio

$$\frac{\Lambda_b^o \to \Lambda_c^+ \rho^-}{\Lambda_b^o \to \Lambda_c^+ \pi^-} \cong 3 \qquad (33)$$

This value differs from the results of ref. [1] probably in the choice of $f_\rho \simeq 0.221$ GeV as compared to $f_\rho \simeq .11$ GeV used in their work but is in line with that of Mannel et al. [18]. For these decays no pole term contributes. Hence experimental value for this branching ratio will give an unambiguous value for the ratio $f_\pi / f_\rho$. It will provide direct test of our results as compared to that of Acker et al. [1].

In order to have an idea about the magnitude of branching ratio, we use the experimental lifetime of $\Lambda_b^o = 1.4 \times 10^{-12}$s [21]:

$$B(\Lambda_b^o \to \Lambda_c^+ \pi^-) \cong 2.115 \times 10^{-3}$$

$$B(\Lambda_b^o \to \Lambda_c^+ K^-) \cong 1.73 \times 10^{-4}$$

$$B(\Lambda_b^o \to \Lambda_c^+ D^-) \cong 2.62 \times 10^{-4}$$

$$B(\Lambda_b^o \to \Lambda_c^+ D_S^-) \cong 7.58 \times 10^{-3}$$

$$B(\Lambda_b^o \to \Lambda_c^+ \rho^-) \cong 6.64 \times 10^{-3}$$

$$B(\Lambda_b^o \to \Lambda_c^+ K^{*-}) \cong 3.4\times10^{-4} \quad (34)$$

Comparing for the branching ratios of the pseudoscalar and vector mesons

$$\frac{B(\Lambda_b^o \to \Lambda_c^+ \pi^-)}{B(\Lambda_b^o \to \Lambda_c^+ K^-)} \cong 12.22$$

$$\frac{B(\Lambda_b^o \to \Lambda_c^+ \rho^-)}{B(\Lambda_b^o \to \Lambda_c^+ K^-)} \cong 19 \quad (35)$$

The branching ratios for vector mesons are very large, probably due to their large masses.A direct measurement of their branching ratio would be apt to discern the overlap of S, P and D-waves in their decays [1]. The experimental data for these decays will become available in the near future. A comparative study of the $\rho^-$ and $D_S^-$ decay will provide further insight into their mechanism and interactions.

It is observed that the complete order-$\alpha_s$ renormalization of the form factors decreases the decay width by about 37% (Table 3) of the value obtained in the leading logarithmic approximation (LLA) in the effective theory. Away from zero recoil point one has to take recourse to full non-perturbative calculations. Consequently, these estimations (Table 3) should be closer to experimental results. The future experimental data will be able to test these predictions. The following points are noted about these results:

1. The value of the decay width is saturated by contributions from the form factors $V_1$ and $A_1$, the others contributing negligibly.
2. The vector and axial-vector components contribute equally to the total amplitude in the HQET limit. However, for renormalization in the full theory, the axial-vector component contributes 2/3 of the vector component to the total amplitude. It is this factor that suppresses the decay width in the full theory.
3. In the real world one expects that $V_1$ should not be equal to $A_1$, since the quarks are not infinitely massive. Only in the HQET limit one has $V_1 = A_1 = \zeta(y)$. An experimental observation of these form factors in $\Lambda_b^o \to \Lambda_c^+ e^- \bar{\nu}$ should provide a direct test for the validity of the HQET.

Similar suppression in the full theory has been pointed out by Neubert [25]; Dosch [26] has also remarked that it is necessary to compute the perturbative QCD renormalization of the form factors beyond the LLA.

In table 4, we compare our results with the works of other people [3-6].For the modes $\Lambda_b^o \to \Lambda_c^+\pi^-$, $K^-$, $D^-$ the values predicted by different authors are comparable. The factorization ansatz should, therefore, work reasonably well for bottom baryons. How the pole model [4] gives a similar results still needs to be understood. The decays $\Lambda_b^o \to \Lambda_c^+ D_S^-, \rho^-$ however, show considerable variation. The small values predicted by Cheng [3] can be attributed to the calculation of the form factor in non-relativistic quark model. Since momentum transfer is large in bottom baryon decays, the form factors should be calculated in relativistic quark model as done by Ivanov et al. [6]. Since the mass of the $\rho$ and D-meson is large, it would be interesting to calculate these decays in the relativistic quark model.

We look forward to confronting our results with more accurate data!

**5. Conclusions**

We have analyzed exclusive $\Lambda_b^o \to \Lambda_c^+ \mathrm{P}^-(V^-)$ non-leptonic decays assuming the factorization approximation added by the current algebra terms. Form factors for renormalization in full QCD are also considered. The decay rates and asymmetry parameters have been calculated. Comparison is also made with predictions of other authors.

It is seen that the pole term and/or ETC may not contribute significantly to the bottom sector. The predicted branching ratios are considerably small when form factors for full QCD are employed. The mode $\Lambda_c^+\rho$, is found to be three times the $\Lambda_c^+\pi$ mode as compared to [1]. An experimental measure of the form factors $V_1$ and $A_1$ in $\Lambda_b^o \to \Lambda_c^+ e^- \bar{\nu}$ is necessitated to test the validity of HQET. Finally, model independent form factors, as in HQET, are desired to have deeper insight into the actual dynamics of these decays.

**Table 1**. Numerical estimates for various terms contributing to the s- and p- wave amplitudes in different processes.

| Decay mode | $A^{fac}$ | $A^{pv}$ | $A^{tot}$ | $B^{fac}$ | $B^{pc}$ | $B^{tot}$ |
|---|---|---|---|---|---|---|
| $\Lambda_b^o \to \Lambda_c^+ \pi^-$ | 1.031 | 0.0 | 1.031 | 2.423 | 0.068 | 2.491 |
| $\Lambda_b^o \to \Lambda_c^+ K^-$ | 0.299 | 0.048 | 0.347 | .705 | -0.149 | 0.556 |
| $\Lambda_b^o \to \Lambda_c^+ D^-$ | 0.411 | 0.0 | 0.411 | 0.967 | 0.0 | 0.967 |
| $\Lambda_b^o \to \Lambda_c^+ D_S^-$ | 2.267 | 0.0 | 2.267 | 5.335 | 0.0 | 5.335 |

**Table 2**. Values of decay width Γ and asymmetric parameter α for renormalization in effective theory assuming the factorization approximation. All widths are in units of $10^{-15}$ GeV.

| Decay mode | Decay width Γ | Asymmetric parameter α |
|---|---|---|
| $\Lambda_b^o \to \Lambda_c^+ \pi^-$ | 3.913 | -1.00 |
| $\Lambda_b^o \to \Lambda_c^+ K^-$ | 0.331 | -0.999 |
| $\Lambda_b^o \to \Lambda_c^+ D^-$ | 0.510 | -0.988 |
| $\Lambda_b^o \to \Lambda_c^+ D_S^-$ | 15.211 | -0.984 |
| $\Lambda_b^o \to \Lambda_c^+ \rho^-$ | 12.883 | - |
| $\Lambda_b^o \to \Lambda_c^+ K^{*-}$ | 0.662 | - |

**Table 3**. Values of decay width Γ in units of $10^{-15}$ GeV for renormalization in full QCD and for factorization approximation.

| Decay mode | QCD effects in full theory |
|---|---|
| $\Lambda_b^o \to \Lambda_c^+ \pi^-$ | 2.602 |
| $\Lambda_b^o \to \Lambda_c^+ K^-$ | 0.214 |
| $\Lambda_b^o \to \Lambda_c^+ D^-$ | 0.323 |
| $\Lambda_b^o \to \Lambda_c^+ D_S^-$ | 9.324 |
| | - |
| $\Lambda_b^o \to \Lambda_c^+ \rho^-$ | 8.1792 |
| $\Lambda_b^o \to \Lambda_c^+ K^{*-}$ | 0.419 |

**Table 4**. The comparison of predicted decay rates Γ (in units of $10^{-15}$ GeV) for various modes computed by different authors and present work.

| Modes→ Authors↓ | | $\Lambda_b^o \to \Lambda_c^+ \pi^-$ | $\Lambda_b^o \to \Lambda_c^+ K^-$ | $\Lambda_b^o \to \Lambda_c^+ D^-$ | $\Lambda_b^o \to \Lambda_c^+ D_S^-$ | $\Lambda_b^o \to \Lambda_c^+ \rho^-$ |
|---|---|---|---|---|---|---|
| Hai-Yang Cheng [3] | | 2.015 | - | - | 6.045 | 2.860 |
| M. R. Khodja et al [5] | | - | - | - | 12.220 | 13.310 |
| S. Sinha et al [4] | | 2.550 | 0.200 | 0.550 | 12.800 | - |
| M. A Ivanov et al [6] | | 2.483 | - | - | - | - |
| Present work | HQET | 3.913 | 0.331 | 0.510 | 15.211 | 12.88 |
| | Full QCD | 2.602 | 0.214 | 0.323 | 9.324 | 8.18 |